\documentclass[twocolumn,showpacs,prb,aps,floatfix,amssymb]{revtex4}

\usepackage{epsfig}
\usepackage{graphics}

\begin{document}

\title{Slow dynamics in a 2D Ising model with competing interactions}  
 
\author{Pablo M. Gleiser}
 \email{gleiser@ffn.ub.es}
 \affiliation{Facultat de F\'{\i}sica, Universitat de Barcelona,
              Barcelona 08028, Spain}
 \author{Francisco A. Tamarit}
 \email{tamarit@famaf.unc.edu.ar}
 \author{Sergio A. Cannas}
 \email{cannas@famaf.unc.edu.ar}
 \affiliation{Facultad de Matem\'atica, Astronom\' {\i}a y F\'{\i}sica,
              Universidad Nacional de C\'ordoba,
              Ciudad Universitaria, 5000 C\'ordoba, Argentina}
\author{Marcelo A. Montemurro}
 \email{mmontemu@ictp.trieste.it}
 \affiliation{Abdus Salam International Centre for Theoretical Physics,
              Strada Costiera 11, 34014 Trieste, Italy}

\date{\today} 
 
\begin{abstract} 
The far--from--equilibrium low--temperature dynamics of ultra-thin magnetic 
films is analyzed by using Monte Carlo numerical simulations on  a two 
dimensional Ising model with competing exchange ($J_0$)  and dipolar ($J_d$) 
interactions.
In particular, we focus our attention on the  low temperature region of the
$(\delta,T)$ phase diagram (where $\delta= J_0/J_d$)  for the range of  values
of $\delta$ where striped phases with widths $h=1$ ($h1$) and $h=2$ ($h2$) are
present. The presence of   metastable states
of the phase $h2$  in the region where the phase $h1$ is the
thermodynamically stable one and viceversa was established recently. In this
work  we show that the presence of these metastable states appears as a
blocking mechanism that slows the dynamics of  magnetic domains growth when
the system is quenched from a high temperature state to a low
temperature state in the region of metastability.  

\end{abstract}

 
\maketitle

\section{\label{intro}INTRODUCTION}
In the last years a strong effort has been devoted to understand both the 
equilibrium and   out-of equilibrium properties of ultrathin magnetic films
\cite{Debell}. These materials have attracted much attention mainly due to their
potential applications, such as information storage \cite{Sampaio}.
Ultra-thin films find also many important applications both in biotechnology
and pharmacology.
It is today a well established experimental fact that the magnetization 
processes 
in ultra-thin magnetic films are ruled by the microscopic competition between
short
range ferromagnetic couplings and long-range frustrated antiferromagnetic
dipolar interactions, which give place to very novel dynamical and static
behaviors \cite{Debell}.

It is worth mentioning that both the theoretical and the experimental interest 
in studying systems with competition
between short-range ordering interactions and long--range frustrating 
interactions
widely exceeds the field of ultra--thin films. Actually, many different 
experimental systems can be  modeled by this kind of microscopic 
interactions, which
give place to very rich dynamical and static properties. In soft--matter
physics for instance, we can mention diblock copolymer melt and cross--linked
polymer mixtures, among others. Type I superconductors and rare--earth layers 
that occur in the 
perovskite structure of REBa$_2$Cu$_3$O$_7-\delta$ (where RE represents a rare
earth from the lanthanide series) can be very well modeled with these
interactions \cite{Macisaac}. 

It has also been frequently suggested that
competing interactions can explain many of 
the phenomenological features observed in the glass formation process and
in supercooled liquids \cite{Kivelson}. Summarizing, many of the conclusions drawn from 
this
work can be surely be applied to a large variety of physical systems.

For sufficiently thin films the
magnetic moments align perpendicular to the plane of the film, indicating that
the surface anisotropy is sufficient to overcome the anisotropy of the dipolar
interaction which favors in-plane ordering.  

Works in two dimensional uniaxial spin systems,
where the spins are oriented perpendicular to the lattice and coupled with
these kind of interactions, have shown a very rich phenomenological  scenario
concerning both its equilibrium statistical mechanics \cite{Kashuba,Macisaac}
and non-equilibrium dynamical properties  \cite{Sampaio,Toloza,Stariolo}. 
In
particular, some of these results \cite{Sampaio,Toloza} showed the existence
of different types of  slow relaxation dynamics when the system is quenched
from  a disordered high temperature configuration to a subcritical
temperature, depending on the relative strengths of the dipolar and exchange 
interactions. 

Under these circumstance one can
use a uniaxial  Ising representation for describing  the magnetic moments
\cite{Allenspach,Allenspach2}.
The ultra--thin film  is then described by the  Hamiltonian 

\begin{equation}
H = -\delta \sum_{ \langle i,j \rangle } \sigma_i \sigma_j + \sum_{(i,j)} \frac{\sigma_i 
\sigma_j}{r^3_{ij}}
\label{Hamilton1}
\end{equation}

\noindent where the spin variable $\sigma_i = \pm 1$ is located at site $i$
of a square lattice, the sum $ \sum_{ \langle i,j \rangle }$ runs over all 
pairs of nearest
neighbor sites and $\sum_{(i,j)}$ runs over all distinct pair of sites of the
lattice; $r_{ij}$ is the distance  (in crystal units) between sites $i$ and
$j$, $\delta$ represents the quotient between the exchange $J_0$ and dipolar
$J_d$ coupling  parameters ($\delta = J_0/J_d$). The energy is measured in
units of $J_d$, which is always assumed to be antiferromagnetic $(J_d > 0)$.
Hence $\delta > 0$ means ferromagnetic exchange coupling.

 We have recently studied in detail  \cite{Gleiser} the low temperature phase 
diagram of this system in the region where the change in the relaxation
properties has been observed. We showed that for very low temperatures metastable
states appear. In this work we investigate the effects of the presence of
these metastable states on the far-from equilibrium dynamical properties of the
system.  In
section \ref{meta} we present a review of the equilibrium phase diagram and
metastability properties in the region of interest. In section \ref{slow} we
analyze the magnetic domain growth or coarsening dynamics of the system  when
it is quenched from  a disordered state (which corresponds to
infinite temperature)  to  a temperature below the ordering transition for
different values of $\delta$. Using Monte Carlo simulations we study the
statistics of domains of the striped phases h1 and h2. We then analyze the
temporal behavior of the average linear size of the domains $L$. We show that
the coarsening dynamics is strongly affected by the presence of metastable
states, which generate blocking clusters of the metastable phase where the
domain walls of the stable phase become pinned.
Such blocking clusters generate free--energy barriers to the domain
growth dynamics that are independent of the linear domain size.
Some conclusions and remarks are summarized in section \ref{conclu}.

\section{Equilibrium phase diagram and metastable states}
\label{meta}

The overall features of the finite temperature phase diagram associated with
Hamiltonian (\ref{Hamilton1}) were described by MacIsaac and coauthors
\cite{Macisaac} by means of Monte Carlo simulations on $16 \times 16$ lattices
and analytic calculations of the ground state \cite{Debell}. They found
that the ground state of Hamiltonian (\ref{Hamilton1})  is the
antiferromagnetic state for $\delta < 0.425$ \cite{Diferences}. For $\delta >
0.425$ the  antiferromagnetic state becomes unstable with respect to the
formation of striped domains structures, that is, to state configurations with
spins aligned along a particular axis forming  ferromagnetic stripes of
constant width $h$, so that spins in  adjacent stripes are anti-aligned,
forming a super lattice in the direction perpendicular to the stripes.
 At high temperatures, of course,  the system always becomes paramagnetic. 
Specific heat calculations showed that the transition between the paramagnetic
 and the striped phases is a second order one \cite{Macisaac}. 

\begin{figure}
\begin{center}
\centerline{\includegraphics[width=0.9\columnwidth]{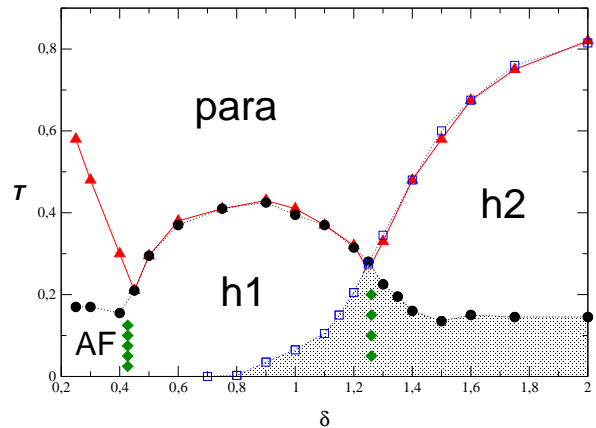} }
\caption{\label{diagrama_fases}Phase diagram $(\delta,T)$ in the region of
parameters under study. Filled triangles correspond to the  critical
temperatures $T_c(\delta)$ obtained by specific heat calculations for the
phase transition between the ordered antiferromagnetic (AF) and striped phases
$h1$  and $h2$ and the paramagnetic ({\bf para}) one. Filled circles (open
squares) correspond to the stability line of the $h1$ ($h2$) phase, obtained by
analyzing the staggered magnetization $M_{h1}$ ($M_{h2}$). Filled diamonds
correspond to the first order transition lines between the $h1$ and $h2$
phases and also between the $h1$ and $AF$ phases, obtained by the free energy
numerical calculations. The shaded region indicates the presence of metastable
states.}  
\end{center}\label{fig1}
\end{figure}

We have recently \cite{Gleiser}  performed Monte Carlo simulations of
Hamiltonian  (\ref{Hamilton1}) on square lattices up 
to $48 \times 48$ sites using periodic boundary
conditions and heat bath dynamics. Our calculations focused on the low 
temperature
 region of the $(\delta,T)$ phase diagram for values
of $\delta$ between $0.2$ and $2$, which includes the transition line between
the  $AF$ and the striped phase with width $h=1$ ($h1$) and  also the
transition line between the $h1$ phase and the striped phase with width $h=2$
($h2$).  First we calculated, through the energy fluctuations, the specific
heat $C$ as a function of temperature for different values of $\delta$ and
different system sizes up to $48 \times 48$ sites.  By considering the peaks
in the specific heat we obtained the second order critical line between the
paramagnetic and the low  temperature ordered phases $h1$ and $h2$. These
results (see Fig. \ref{diagrama_fases}) slightly improved those obtained by
MacIsaac  and coauthors \cite{Macisaac} for $16 \times 16$ lattices, thus
showing a fast convergence of the critical temperature for increasing system
sizes, at least for small values of $\delta$.

Next, through a numerical study of the free energy, we analyzed the transition
 between the $h1$ and  $h2$ phases.  The free energy of the phase $h1$ ($h2$)
was  calculated for increasing (decreasing) values of $\delta$. We observed a
continuous  change of the minimal free energy from one phase  to the other,
with a discontinuous change in the slope for  $\delta = 1.26(1)$ indicating
the presence of a first order phase transition \cite{Gleiser}. This transition
is indicated by means of diamonds in Fig.  \ref{diagrama_fases}.  We have also
repeated these calculations for the transition line between the $h1$ and $AF$
phase, finding similar results.

Close to $\delta = 1.26(1)$ the free energy displays a multivalued behavior 
characteristic of a first order phase transition. This 
behavior signals the metastable nature of these phases in some parts of the 
phase diagram.  To characterize the presence of metastable states observed in
the transition between the $h1$ and $h2$ phases we  introduced the staggered
magnetizations  $M_{h1}$ and $M_{h2}$, and also their associated
susceptibilities $\chi_{h1}$ and $\chi_{h2}$ for the $h1$ and $h2$ phases 
\cite{Gleiser}.
These quantities permitted us to analyze the stability of both phases in the
different parts of the  phase diagram. In Fig. \ref{diagrama_fases} the shaded
region indicates the presence of metastable states. It is important to stress
that  in the shaded region inside phase $h1$ the only phase observed to be
metastable was $h2$. For $\delta > 1.26$ metastables states of phases of
higher width were observed, leading to a much  more complicated metastable
region. In this work we will focus only on the  dynamical behavior of the
system in the $h1$ region. As we will show in the next section, for a fixed
value of $\delta$ inside  this region different dynamical regimes are
observed  as the temperature is lowered and one enters the
 region of metastability.

\section{SLOW DYNAMICS}
\label{slow}

When a system is quenched from a high temperature disordered phase  into a
low temperature ordered phase domains form and grow, a process that is known
as coarsening. The coarsening process has been extensively studied both
experimentally and theoretically over the past  decade \cite{Bray}.

Perhaps the most thoroughly studied system, and also the most common example
is  the Ising model. When this system is quenched from  a high temperature to
one below its critical temperature ($T<T_c$) ferromagnetic domains of up and
down spins form and coarsen. The system presents curvature driven growth and
the characteristic domain size $L$ grows with time as $L(t) \sim t^{1/2}$. If
the system is cooled to zero temperature the domain walls can be easily
determined as bonds between oppositely oriented spins, but if the system is
cooled to  a temperature different from zero it becomes difficult to define
domains and domain walls since small islands generated by thermal fluctuations
arise.  To overcome this problem, Derrida \cite{Derrida} proposed a new method
to measure properties related to  coarsening in the presence of thermal
fluctuations. This method was extended by Hinrichsen and Antoni
\cite{Hinrichsen2} to determine domain walls for nonzero temperatures. The
method compares the state of a system with replicas in the different ground
state  configurations when they are all submitted to the same thermal noise,
that is, when the same sequence of random numbers is used to update all
systems. In this way, if one starts  from a replica in the ordered state a spin flip 
will be a consequence of the thermal noise. When a spin flip occurs simultaneously in 
all the replicas it can be considered as a thermal fluctuation, otherwise the fluctuation 
will be due to the coarsening process. We used this technique to study the dynamics 
of domain walls, and characterize the coarsening process, when the system described  
by Hamiltonian (\ref{Hamilton1}) is quenched from a high temperature disordered state 
into the region where it orders. In particular we  focus our interest on the growth 
of domains of the striped phase $h1$ when the metastability line is crossed for values 
of $0.8 < \delta<1.26$ (See Fig.\ref{diagrama_fases}).

To characterize the growth of the domains we determined first the domain areas
 $A(t)$, by counting the number of spins inside each domain.  The
characteristic (linear) domain size was calculated as $L(t) = \sqrt{\langle
A(t) \rangle}$, where  $\langle A(t) \rangle$ is the mean domain area of the
system at time $t$.

\begin{figure}
\begin{center}
\centerline{\includegraphics[width=0.9\columnwidth]{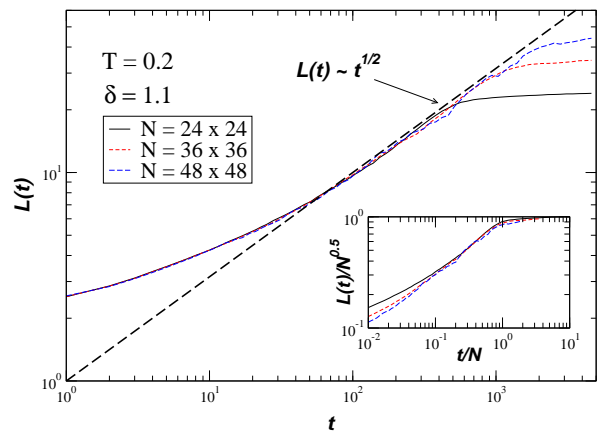} }
\caption{\label{dinamica1}Characteristic domain size $L(t)$ vs. $t$ when
$\delta = 1.1$ and $T=0.2$ for three different system
 sizes $N = 24 \times 24$,  $N = 36 \times 36$ and $N = 48 \times 48$ .  The
dashed line indicates $L(t) \sim t^{1/2}$. The inset shows the data collapse
obtained using finite size scaling analysis.}
\end{center}
\end{figure}

In Fig. \ref{dinamica1} we present the behavior of the characteristic domain
size $L(t)$ when $\delta = 1.1$ and $T=0.2$ for  three different system sizes
$N = 24 \times 24$,  $N = 36 \times 36$ and $N = 48 \times 48$. After a short
transient in which the  characteristic  length presents a slow growth the
system enters into a coarsening regime where $L(t) \sim t^{1/2}$, as expected
for a system with non--conserved order parameter\cite{Bray}. For large times
$L(t)$ presents a crossover to a saturation value.  This saturation behavior
is  clearly a finite size effect, since a domain cannot grow beyond the system
size $N$ so that as the system size increase,  the behavior $L(t) \sim
t^{1/2}$ remains for larger periods of time.

\begin{figure}
\begin{center}
\centerline{\includegraphics[width=0.9\columnwidth]{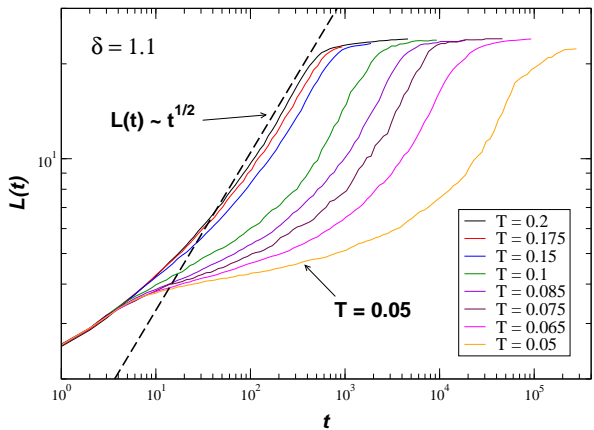} }
\caption{\label{dinamica2}Characteristic domain size $L(t)$ vs. $t$ when
$\delta = 1.1$ ($N=24 \times 24$) for eight different temperatures,  starting 
form the left
curve $T=2.0$, $0.175$, $0.15$, $0.1$, $0.085$, $0.075$, $0.065$  and
$T=0.05$. Note that all the  curves eventually cross-over to the $L(t) \sim
t^{1/2}$ regime indicated with a dashed line.} 
\end{center}
\end{figure}

In section \ref{meta}  we observed and characterized the presence of metastable 
states in the low temperature region of the 
phase diagram. We will study now how the behavior of $L(t)$  changes as we lower
the temperature and cross the metastability line
for a fixed value of $\delta$. In Fig. \ref{dinamica2}
we present the time evolution of $L(t)$ when  $\delta = 1.1$ for eight 
decreasing temperatures. For low temperatures a regime of slow 
growth develops at intermediate time scales before 
the system crosses over to the $t^{1/2}$ coarsening regime. As we lower the 
temperature the intermediate regime extends to larger
 time scales,  but all the curves eventually cross over to the $L(t) \sim 
t^{1/2}$ regime. Note that for short times there seems
 to be a change in the concavity of $L(t)$ when it crosses $T=0.1$, which 
coincides with the boundary
 of the metastable phase for $\delta = 1.1$.

\begin{figure}
\begin{center}
\includegraphics[width=0.85\columnwidth,angle=-90]{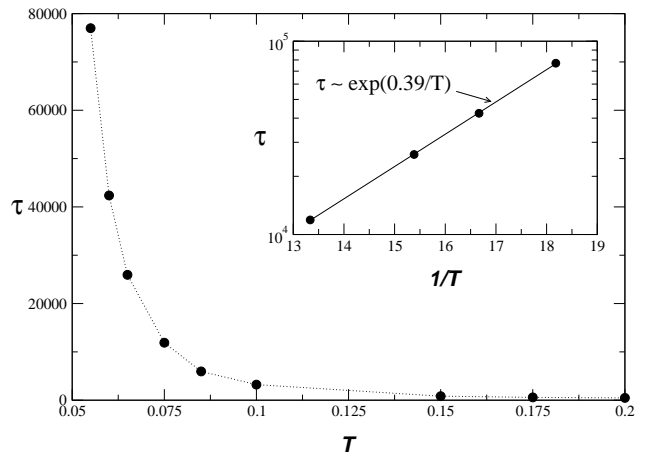}
\caption{\label{dinamica3}Crossover time $\tau$ vs $T$ for $\delta=1.1$ and 
$N=24 \times 24$. On the inset an
Arrhenius plot $\tau$ vs $1/T$ for the four lower temperatures
 is presented. The straight line indicates the best fit.}
\end{center}
\end{figure}

To characterize the behavior of $L(t)$ as we lower the temperature we 
studied the crossover time $\tau$ from the power law to the saturation regime. 
Note that the crossover time to the power law regime presents a similar 
behavior.  However the study of the intersection of the power law branch
of the curve and the horizontal saturation branch allows for a sistematic
approach. This is so since the saturation value always correspond to 
the linear size of the system. In figure
\ref{dinamica3} we present how $\tau$ grows as  the temperature is lowered.
For temperatures greater than $T=0.1$ the crossover time presents a linear
dependency with $1/T$, while for temperatures lower than $T=0.1$ it  presents
an exponential increase with $1/T$ as can be observed in the Arrhenius plot
presented in the inset of Fig. \ref{dinamica3}. 

\begin{figure}
\begin{center}
\centerline{\includegraphics[width=0.9\columnwidth]{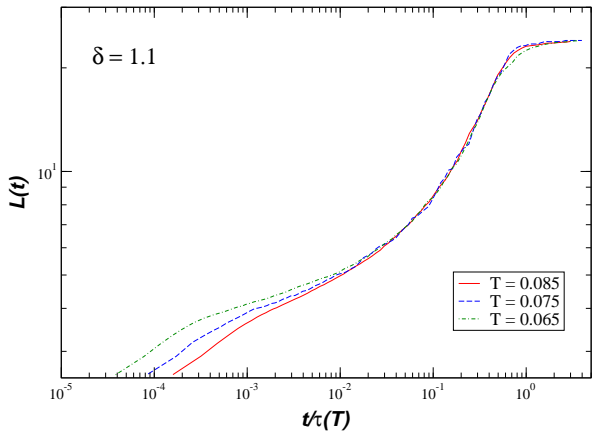} }
\caption{\label{dinamica4}Data collapse plot of $L(t)$ in the low temperature
regime for three different temperatures ($\delta=1.1$ and $N=24 \times 24$). The 
crossover time $\tau$ corresponds
to the
best fit  presented in Eq. (\ref{eqtau}).}
\end{center}
\end{figure}

The straight line indicates the best fit, given by a function of the form
\begin{equation}
\tau = \tau_0 \exp(\tau_1/T)
\label{eqtau}
\end{equation}
where $\tau_0=62.5(5)$ and $\tau_1=0.39(5)$. Using this expression we present
 in Fig. \ref{dinamica4} a data collapse  plot of $L(t)$ in the low 
temperature regime. 

\begin{figure}
\begin{center}
\centerline{\includegraphics[width=0.85\columnwidth]{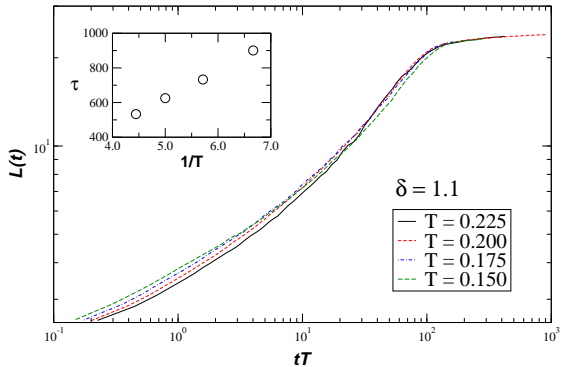} }
\caption{\label{dinamica5}Data collapse plot of $L(t)$ in the high temperature
regime for three different temperatures $\delta=1.1$ and $N=24 \times 24$.}
\end{center}
\end{figure}

In the high temperature regime the crossover time decreases linearly as the 
temperature increases, and $L(t)$ collapses simply
by scaling with $T$ as can be seen in figure \ref{dinamica5}.

Summarizing, for $\delta = 1.1$ two different dynamical regimes were observed
above and below the metastability line ($T=0.1$).
When the system is quenched to the ordered phase
to a temperature $T>0.1$, the characteristic domain size $L(t)$ grows 
as $t^{1/2}$ after a short transient. If, on the other hand,  the system 
is quenched to a  temperature $T<0.1$,  an intermediate regime with a
 slow growth  appears. For long times the system always crosses over to 
the $t^{1/2}$ regime. That is, for every temperature we observe the same
 behavior presented  in Fig.\ref{dinamica2}, where $L(t)$ always reaches the 
asymptotic  behavior $t^{1/2}$ as we increase the system size. However, as we
 lower the  temperature the crossover time to this regime increases. We 
repeated these analysis for different values of $\delta$, inside the $h=1$ 
region, obtaining  the same qualitative behaviors.

These dynamical behaviors present a strong resemblance with the ones observed
 in the two dimensional Shore model \cite{Shore2}. This model
is a ferromagnetic Ising model on a square lattice with frustration added by
introducing weak next-nearest-neighbor antiferromagnetic
 bonds, that is

\begin{equation}
 H = -J_1 \sum_{NN} s_i s_j + \sum_{NNN} s_i s_j
\end{equation}

The presence of NNN antiferromagnetic bonds in this model introduce free--energy
barriers to domain coarsening that are independent of the domain size
$L$\cite{Shore2}.
Such barriers in this model are a consequence of a corner rounding
process which generates structures that block the coarsening
dynamics\cite{Shore2}.
Hence, the system is stuck and coarsens little on time 
scales $t \ll \tau_B(T) = exp{(F_B /T)}$ ($F_B$ being the height of
 the barrier), while on time scales $t \gg \tau_B(T)$ the free--energy barrier
 can be crossed and the $t^{1/2}$ behavior emerges.

\begin{figure*}
\centerline{\includegraphics[width=14cm]{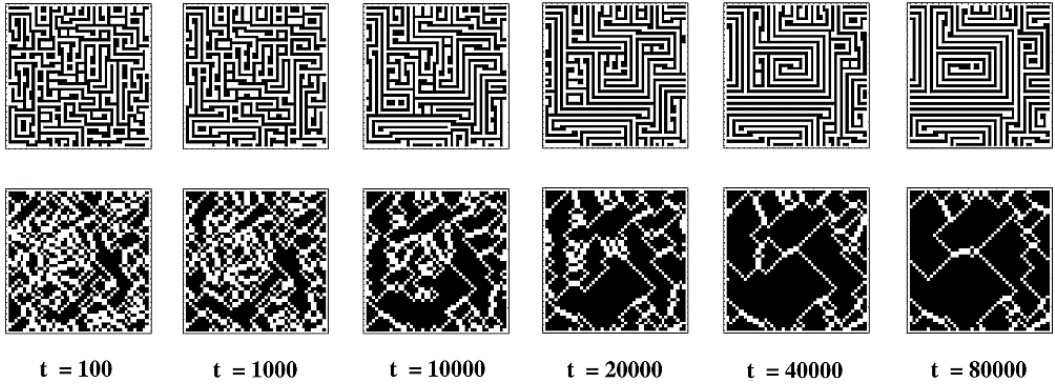} }
\caption{\label{dinamica6} Snapshots of the coarsening process for
$\delta = 1.1$, $N=48 \times 48$ and $T=0.05$. The squares at the top correspond
to spin configurations at different times of the coarsening (black points: up
spins; white points: down spins). The
black areas in the figures at the bottom are the corresponding domains of the
$h1$ phase at every time, while the white lines correspond to the domain walls.}
\end{figure*}

\begin{figure}
\includegraphics[width=0.65\columnwidth,angle=-90]{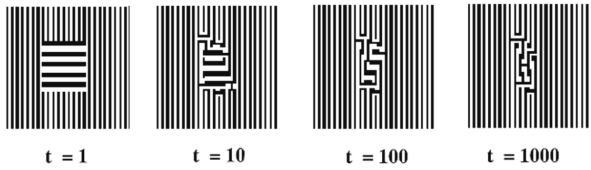}
\caption{\label{square} Snapshots of the coarsening process for
a $4 \times 4$ shrinking square in phase $h2$ immersed in a system in phase $h1$
 with fixed boundary conditions.}
\end{figure}

\begin{figure}
\includegraphics[width=0.8\columnwidth,angle=-90]{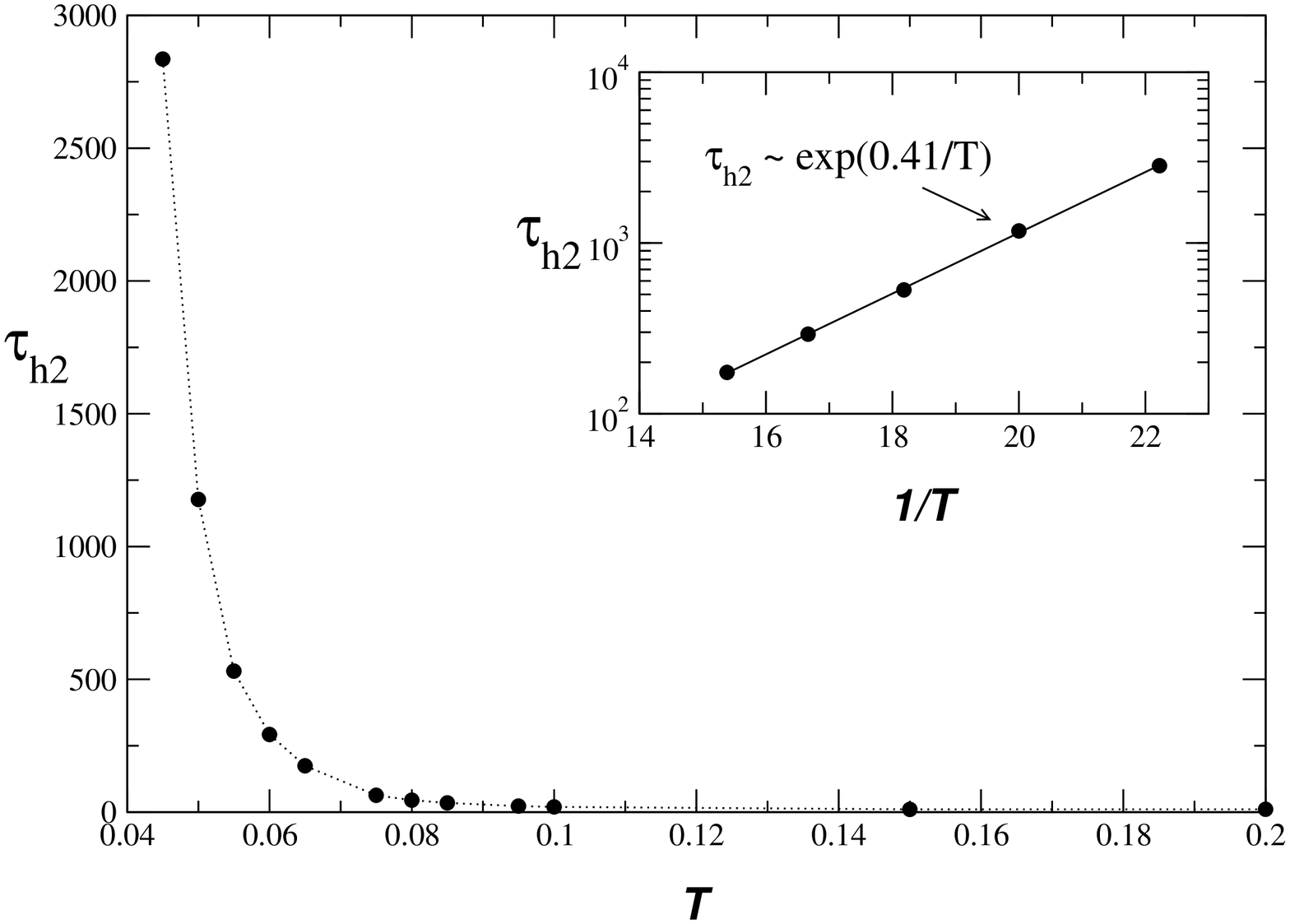}
\caption{\label{dinamica7} Shrinking time of a $4 \times 4$ square in phase $h2$
immersed in a system in phase $h1$.}
\end{figure}

Thus, it is interesting to see if structures that block the coarsening process
are present in our system. In fact the presence
of metastable states for low temperatures gives a strong hint of what to look
for.

Figure \ref{dinamica6} presents a series of snapshots of the coarsening process when 
a system of size $N = 48 \times 48$ and $\delta = 1.1$ is quenched to $T=0.05$.
For these values of $\delta$ and $T$ the system evolves in the metastability region. 
Two different kind of domain walls can be clearly distinguished. When domains of phase 
$h1$ with different orientation meet, the walls detected by the method are thin. The 
thicker walls correspond to bands of two spins, that is stripes of  phase h2. We should 
stress that phase h2 is the only metastable phase in phase h1 \cite{Gleiser}.  
Notice how these small  blocks of the h2 phase seem to slow the domain growth by pinning 
the domain walls.  These small blocks were also observed at short times when the system 
was quenched to a temperature above the  metastability line,  however they are highly 
unstable and  did not seem to block the domain growth. 
In order to
quantify this effect, we studied the time for completely
shrinking squares in phase h2 immersed in a system in phase h1
with fixed boundary conditions. A square was considered to be completely
shrunk when the length of the sorrounding domain walls became zero.
Fig.\ref{square} presents a series of snapshots of this process. In the first time steps the square quickly deforms. However, further 
advance seems to be blocked by small blocks of spins in the $h2$ phase. 
This coarsening behavior presents a strong resemblance to the one 
observed in the  snapshots presented in Fig. \ref{dinamica6}.

Figure \ref{dinamica7} shows the time $\tau_{h2}$  to shrink a $4
\times 4$ square in phase $h2$ immersed in  a system  with $N = 48 \times 48$
spins in phase $h1$ with fixed boundary conditions. As the temperature is
lowered the shrinking time diverges as $\tau_{h2} \sim exp (0.41/T)$. This
divergence agrees well with the divergence observed in the crossover time  from
the  slow growth to the $t^{1/2}$ regime.  We have also studied the shrinking 
time of blocks in different phases,
such as combinations of vertical and horizontal $h1$ phases and also a
ferromagnetic block immersed in the $h1$ phase. In all these cases a
similar behavior was observed.
The system quickly reached a configuration where small blocks of the $h2$
phase
were present, with a slowing down of the domain growth similar to the one
described above.

\vspace{0.5cm}

\section{Conclusions}

\label{conclu}
In this paper we have studied some
dynamical properties of a two-dimensional Ising Hamiltonian
with competing interactions. 
We analyzed the coarsening process when the system
is quenched from a high temperature disordered phase
into the ordered phase $h1$. To characterize the
growth of domains we considered the time evolution
of the characteristic domain size $L(t)$. We found
that for a fixed value of $\delta$ the system presents
two different dynamical behaviors associated with the
presence or absence of $h2$ metastable states. When the
system is quenched into the ordered phase for 
temperatures above the metastability region, the characteristic
domain length presents a power law $t^{1/2}$ growth.
If, on the other hand, the system is quenched to 
temperature in the metastability region, the behavior
of $L(t)$ presents a slow growth intermediate regime
before crossing over to the $t^{1/2}$ power law growth.
Through a direct examination
of snapshots of the system during the coarsening
process we found that the presence of small domains
in the $h2$ phase slowed the coarsening when the
system was quenched to the metastable region.
These results are consistent with the presence of
free--energy barriers independent of the domain size $L$,
associated with blocking clusters of the metastable phase,
which generates a crossover in the coarsening behavior as
we cross the spinodal line. Since the cross over time diverges
as the temperature is lowered, the very slow behavior at intermediate
times may be indistinguishable from a logarithmic law. This
could explain the apparently logarithmic scaling observed
in the aging behavior of this model\cite{Toloza}.

This work was partially supported by grants from
Consejo Nacional de Investigaciones Cient\'\i ficas y T\'ecnicas CONICET  
(Argentina), Consejo Provincial de Investigaciones Cient\'\i ficas y  
Tecnol\'ogicas (C\'ordoba, Argentina),  Secretar\'\i a de Ciencia y  
Tecnolog\'\i a de la Universidad Nacional de C\'ordoba (Argentina) and 
Fundaci\'on Antorchas (Argentina).

\bibliographystyle{prsty}

\begin{thebibliography}{}
%
\bibitem{Debell} K. De'Bell, A. B. MacIsaac and J. P. Whitehead, Rev. Mod. Phys.
{\bf 72}, 225 (2000).
%
\bibitem{Sampaio} L. C. Sampaio, M. P. de Albuquerque and F. S. de Menezes,
 Phys. Rev. B {\bf 54}, 6465 (1996).
%
\bibitem{Macisaac} A. B. MacIsaac, J. P. Whitehead, M. C. Robinson and  K.
De'Bell,
 Phys. Rev. B {\bf 51}, 16033 (1995).
%
\bibitem {Kivelson} D. Kivelson, S. A. Kivelson, X. Zhao, Z. Nussinov and T. Gilles. 
Physica A {\bf 219}, 27 (1995).
%
\bibitem{Kashuba} A. Kashuba and V. L. Pokrovsky.  Phys. Rev. Lett. {\bf 70},
3155 (1993).
%
\bibitem{Toloza} J. H. Toloza, F. A. Tamarit and S. A. Cannas,
Phys. Rev. B {\bf 58}, R8885 (1998).
%
\bibitem{Stariolo} D. A. Stariolo and S. A. Cannas Phys. Rev. B {\bf 60}, 3013 
(1999).
%
\bibitem{Allenspach} R. Allenspach, M. Stampanoni and A. Bischof,
Phys. Rev. Lett. {\bf 65}, 3344 (1990).
%
\bibitem{Allenspach2} R. Allenspach, and  A. Bischof,
Phys. Rev. Lett. {\bf 69}, 3385 (1992).
%
\bibitem{Gleiser} P. M. Gleiser, F. A. Tamarit and S. A. Cannas, Physica D
{\bf 168--169}, 73 (2002). 
%
\bibitem{Diferences} It is worth noting that MacIsaac and coauthors 
\cite{Macisaac} definition of the Hamiltonian (Eq. (5) in that
reference) is slightly different from ours (Eq(\ref{Hamilton1})). While in that
paper the dipolar term contains a sum over all pairs of spins, in  Hamiltonian
(\ref{Hamilton1}) we consider the sum over every pair of spins just once. This
leads to the equivalence $\delta = J/2$, $J$ being the exchange parameter in
the above reference. Since the dipolar parameter also fixes the energy units in
our work, there is also a factor $1/2$ between the critical temperatures
obtained in both works. 
%


\bibitem{Bray} A. J. Bray,  Advances in Physics {\bf 43}, 357 (1994). 
%

\bibitem{Derrida} B. Derrida,  Phys. Rev. E {\bf 55}, 3705 (1997).
%


\bibitem{Hinrichsen2} Haye Hinrichsen and M. Antoni,  Phys. Rev. E {\bf 57}, 
2650-2655 (1998).

\bibitem{Shore2} J. D. Shore, M. Holzer and J. P. Sethna,  Phys. Rev. B
{\bf 46}, 11376 (1992).


\end{thebibliography}

\end{document}